\documentclass[10pt,letterpaper]{article}
\usepackage[top=0.85in,left=2.75in,footskip=0.75in]{geometry}
\usepackage[T1]{fontenc}

\usepackage{amsmath,amssymb}

\usepackage{changepage}

\usepackage{textcomp,marvosym}

\usepackage{cite}
\usepackage{graphicx}
\usepackage{booktabs}
\usepackage{arydshln}
\usepackage{nameref,hyperref}
\usepackage{hhline}
\usepackage{arydshln}
\usepackage{rotating}
\usepackage{diagbox}
\usepackage{booktabs}
\usepackage{subcaption}

\usepackage[nopatch=eqnum]{microtype}
\DisableLigatures[f]{encoding = *, family = * }

\usepackage[table]{xcolor}
\definecolor{highlight}{RGB}{144, 238, 144}  

\usepackage{array}

\newcolumntype{+}{!{\vrule width 2pt}}

\newlength\savedwidth

\raggedright
\usepackage[aboveskip=1pt,labelfont=bf,labelsep=period,justification=raggedright,singlelinecheck=off]{caption}

\usepackage{float}

\makeatletter
\renewcommand{\@biblabel}[1]{\quad#1.}
\makeatother

\usepackage{lastpage,fancyhdr,graphicx}
\usepackage{epstopdf}
\usepackage{graphicx}
\usepackage{multirow}
\fancyheadoffset[L]{2.25in}
\fancyfootoffset[L]{2.25in}
\begin{document}
\vspace*{0.2in}

\begin{flushleft}
{\Large
\textbf\newline{Urban delineation through the lens of commute networks: Leveraging graph embeddings to distinguish socioeconomic groups in cities} 
}

\bigskip

Devashish Khulbe\textsuperscript{1*},
Stanislav Sobolevsky\textsuperscript{1, 2, 3}
\\
\bigskip
\textbf{1} Department of Mathematics and Statistics, Faculty of Science, Masaryk University, Brno, Czech Republic
\\
\textbf{2} Center for Urban Science+Progress, New York University, Brooklyn, New York, U.S.A
\\
\textbf{3} Center for Interacting Urban Networks, New York University Abu Dhabi, Abu Dhabi, U.A.E
\\
\bigskip

* Corresponding author \\
Email: dk3596@nyu.edu (DK)

\end{flushleft}

\newpage

\section*{Abstract}
Delineating areas within metropolitan regions stands as an important focus among urban researchers, shedding light on the urban perimeters shaped by evolving population dynamics.
Applications to urban science are numerous, from facilitating comparisons between delineated districts and administrative divisions to informing policymakers of the shifting economic and labor landscapes. 
In this study, we propose using commute networks sourced from the census for the purpose of urban delineation, by modeling them with a Graph Neural Network (GNN) architecture. We derive low-dimensional representations of granular urban areas (nodes) using GNNs. Subsequently, nodes' embeddings are clustered to identify spatially cohesive communities in urban areas.  Our experiments across the U.S. demonstrate the effectiveness of network embeddings in capturing significant socioeconomic disparities between communities in various cities, particularly in factors such as median household income. The role of census mobility data in regional delineation is also noted, and we establish the utility of GNNs in urban community detection, as a powerful alternative to existing methods in this domain. The results offer insights into the wider effects of commute networks and their use in building meaningful representations of urban regions.


\section*{Introduction}
Urban delineation plays a crucial role in addressing a wide range of questions in urban science. City administrators rely on well-defined boundaries to allocate resources effectively and plan essential services, while urban scientists and researchers analyze evolving urban boundaries to gain insights into shifting demographic patterns, spatial dynamics, and the changing nature of urban environments.
Recent research has concentrated on discerning business hubs, regional administrative precincts, and even delineations of entire metropolitan areas. 

Dynamic network data has long been proposed by researchers for this purpose. Cell-phone records have proved quite effective in regional delineation, modeling on GPS trajectories and origin-destination movement networks \cite{dong2024defining, li_cellPhone}, while others have also considered urban movement derived from social media interactions \cite{social_media}, where larger regions have also been delineated on country level \cite{ratti_nation_network}. However, a data source that is widely available in cities is not yet widely known. We propose using mobility networks from Origin-Destination commute data available from the census. Such a dataset offers comprehensive coverage, facilitating studies throughout the country, and eliminating the reliance on alternative, potentially biased mobility data providers that may exclude significant portions of the population. Census data boasts extensive spatial reach, enables sophisticated network analysis, and constitutes a vast repository covering the majority of the country's inhabitants.

Network partitioning has been tackled with methods such as community detection \cite{FORTUNATOcommunitydetection}, kernel density estimation \cite{kde_delineation}, and partitioning based on spatial constraints \cite{social_media}, which are generally based on optimizing a network metric such as modularity. While self-supervised deep learning based network embeddings have not been evaluated for detecting communities, they have been widely successful in many downstream tasks with urban networks. With mobility networks, GNN based modeling have shown to be successful in socioeconomic modeling \cite{khulbe2023mobility, khulbe2025commutenetworkssignatureurban}, while various GNN model architectures have been employed for numerous tasks like edge prediction and community detection in urban graphs \cite{liu2024community, sobolevsky2022GNNS}. They have also been useful in heterogeneous urban networks like population-facilities interaction graphs \cite{GAT_population_facilities}. Using census data-derived commute networks as input, we train a GNN model to obtain low-dimensional embedding vectors of census tracts (nodes) in urban networks. The model is learned in a self-supervised fashion with the objective to reconstruct the original commute flows (edges) in the network. To obtain communities, we further cluster the embedding vectors and notice spatially homogeneous communities in all metro areas. Next, we evaluate the socioeconomic profiles of the clusters, and observe varying differences in income status among communities within cities. The income status is directly related to commute ability and flexibility, which is revealed by the changing community structure.


While GNN-based embeddings tend to yield more homogeneous and coherent community structures, graph clustering and community detection have long been active areas of research. Notably, graph embeddings have also been proposed as a powerful approach for community detection~\cite{graphEmbeddingsCD}, often demonstrating performance comparable to traditional methods. However, such techniques have rarely been evaluated in the context of urban mobility networks, where the spatial and socioeconomic dimensions introduce additional complexity. Furthermore, it remains important to assess these methods across real-world networks of varying scales to understand their generalizability and robustness.
Also, it is crucial to have some sort of comparative analysis to existing methods, perhaps with a network-based metric commonly used for community detection evaluation purposes. Therefore, we also compare our results with two widely used community detection approaches--1. A network modularity optimization-based method, and 2. A probabilistic generative Stochastic Block Model.

Our main findings can be summarized in two key points:
\begin{enumerate}
    \item We show the effectiveness of general-purpose GNN-based network embeddings in enabling efficient and socioeconomically meaningful urban delineation, and find them to be at least as good, and in some cities, to be better than traditional community detection methods.
    \item We demonstrate the practical value of census-derived mobility data for urban community delineation, highlighting its accessibility and relevance in diverse geographic contexts.
\end{enumerate}

We present our results for the 12 largest U.S. metropolitan areas, with varying numbers of delineations for each area. We compare the results with community detection methods, and also present a qualitative evaluation of differences in communities from a socioeconomic perspective. Notably, we observe delineation of high-income density neighborhoods from low-income ones. The distinction is interesting, as it reveals commute patterns and gaps, which may be exacerbated by commuters' income and other socioeconomic factors. 

\section*{Data overview}
Diverse data have been proposed for the purpose of urban and regional delineation. 
Mobile phone data has been used in many recent studies to forge mobility networks for various downstream tasks \cite{yuriBrnoMobile}. Mobile phone data is good for highly granular analysis, but not consistently available everywhere. Other studies have also frequently used the social media footprint of people across urban areas and POIs \cite{social_media}. In general, mobility datasets have been widely used for modeling and as a key feature for urban modeling, such as socioeconomic analysis and regional delineation\cite{mobilitySocioeconomic, ratti_nation_network}. Phone-based mobility and social media data may provide more dynamic records, but comes with the condition of not being comprehensive across locations and populations. In this study, we thus propose census-derived mobility data for networks in cities. Census data has long been a valuable resource for researchers, widely utilized in studies ranging from population dynamics to urban sprawl, land use, and development \cite{censusData, censusData1}. Although census-derived mobility data remains relatively underexplored, prior research has demonstrated the effectiveness of census-based variables—such as income and unemployment rates—as indicators of socioeconomic status \cite{censusSocioeconomic, censusSocioeconomic1}.

For the U.S. cities considered in our analysis, we retrieve the Origin-Destination (O-D) commute data from the Longitudinal Employer-Household Dynamics (LEHD) \cite{lehd2025}. Leveraging administrative records from the U.S. Census Bureau, LEHD provides a comprehensive view of worker flows in all cities across the U.S., and thus offers rich insights into the dynamics of labor markets and commuting patterns. Populating the network with LEHD commute flows hence provides a comprehensive picture of mobility across cities. We also use U.S. census data to retrieve income data for all cities considered. Median income is considered a proxy to evaluate the socioeconomic profile of areas in cities.

\begin{table}[!h]
  \caption{Commute network statistics for 12 U.S. cities}
  \label{tab:city_stats}
  \begin{tabular}{lccc} 
    \textbf{City} & \textbf{Nodes} & \textbf{Non-zero Edges} & \textbf{Avg. edge weight} \\
    New York & 2157 & 976832 & 0.69 \\
    Chicago & 1318 & 439553 & 1.06 \\
    Boston & 520 & 127357 & 3.5 \\
    Austin & 218 & 34777 & 8.63 \\
    Dallas & 529 & 129352 & 2.83 \\
    Los Angeles & 2341 & 1171362 & 0.65 \\
    San Antonio & 366 & 83192 & 4.77 \\
    San Diego & 627 & 180781 & 2.97 \\
    San Jose & 372 & 81938 & 4.73 \\
    Philadelphia & 384 & 68119 & 2.57 \\
    Phoenix & 916 & 349894 & 2.10 \\
    Houston & 786 & 290496 & 2.50 \\
  \end{tabular}
\end{table}

Table\ref{tab:city_stats} shows the population and commute network statistics for 12 U.S. metro regions considered in the experiments. The biggest metro areas in the country spans across every geography and covers a total of approximately 30 million people. Using census data here is key, as it is consistently available across the nation spanning across every spatial granularity. This makes our experiments possible across every major urban area in the country. Thus, the accessibility and comprehensiveness of the mobility information from the census is hard to match by other sources.

\section*{Methods}
The underlying idea behind delineation in urban networks is of finding similar nodes in graphs based on network properties. There have been many methods proposed in the literature for this purpose. For comparison with our methods, we will primarily look into the results from community detection methods, which have been widely successful in urban networks in recent years \cite{Sobolevsky2013, Expert2011}.

Our approach mainly work by deriving the embedding of mobility networks. Mobility is a crucial urban metric and is the core ability under which many lifestyle choices depend of people.
Some recent studies have also found that mobility is vastly impacted by commuters' income status \cite{IncomeCommute1, IncomeCommute2}. Thus, embedding as a low-dimensional representation of the larger mobility network can be particularly useful for modeling socioeconomic indicators of regions, including income.

Our methods derive the idea from using positional encoding in models, with recent models with graph modeling with transformer-based models \cite{GT2}. We mainly propose a self-supervised learning framework, with GNN being the model architecture for network information propagation. We also present a MLP-based model, as a deep-learning baseline model to compare how GNN-based embedding compares to a standard MLP-based representation learning methodology. Thus, we first derive low-dimensional representation vectors (embedding) for the commute network i.e. $N \times d$ matrix for N spatial units (nodes), d being the embedding dimensionality.
Then we cluster the embedding vectors using the K-means method to get communities in the urban area.

\subsection*{Positional and Structural encodings (PSE)}

We derive the utility of encoding methods to capture positional and structural information in networks. Laplacian Eigenvectors (LE) has been used as positional encodings of nodes in recent models on Graph Transformers (GT) \cite{GraphTransformers}. Other methods include shortest-path distances \cite{ShortestPathEncoding}, and SVD \cite{svd}, which has also been used in Graph Representation learning tasks \cite{svd_grl}. SVD graph embedding has also proven to be useful in community detection, with promising results for unipartite and bipartite graphs \cite{svdcommunity}. Structural encodings have been developed to capture rich local and global connectivity patterns in networks. Random Walk encoding \cite{RandWalkEncoding0}, has proven to be quite useful in capturing structural information in graphs \cite{encoding3}. They also have been successfully used as structural encoding in GNN based models \cite{RandWalkEncoding1, RandWalkEncoding2}.
Recent works on graph-based models (GNNs and Graph Transformers) use many of these existing embedding methods as positional inputs to the models. In urban mobility networks, network encodings have been proven to be useful for downstream tasks using heterogeneous networks \cite{UrbanEmbedding1, UrbanEmbedding2}.  

Hence, the underlying idea behind experimenting with PSE is the evaluation of delineation capabilities of the encodings, and whether the resulting urban communities are spatially homogeneous. The presumption with using PSE is also that any modeling involves significantly larger parameters and training, whereas PSE are relatively easier and trivial to obtain. Additionally, we also evaluate the capabilities of PSE in discerning neighborhoods based on their socioeconomic profiles.

\subsection*{Representation Learning}

Graph representation learning aims to generate low-dimensional embeddings for each node (or sub-region in our context), from which the original graph structure can be reconstructed or downstream tasks (e.g., community detection, node classification) can be performed. Deep neural networks, particularly GNNs, have been highly successful in capturing meaningful graph representations by leveraging graph connectivity and neighborhood structure. Below, we outline our approaches for node embedding---focusing on our GNN-based method---along with a brief overview of a baseline MLP-based approach.

\subsubsection*{GNN-Based Embedding}

Graph Neural Networks (GNNs) extend conventional neural networks to graph-structured data, enabling node (or sub-region) embeddings to be learned by aggregating and transforming information from local neighborhoods. In our setup, we utilize a two-layer Graph Neural Network (GNN) architecture, which has been shown to be effective for representation learning across a wide range of graph-related tasks~\cite{GNN, GNN_2layer}. Two-layer GNNs are particularly well-suited for capturing local structural information while maintaining computational efficiency. Moreover, increasing the number of layers in GNNs often leads to the oversmoothing phenomenon, where node representations become indistinguishable, ultimately degrading model performance~\cite{GNN2layer, Graph2layerModels}.

\paragraph{Architecture.}
A two-layer GNN can be viewed as:
\begin{equation}
    H^{l} = \sigma \bigl(W^{l} \,\hat{A}\,H^{l-1} + B^{l}\bigr),
\end{equation}
where:
\begin{itemize}
    \item $H^{l-1}$ is the node embedding matrix at layer $(l-1)$,
    \item $\hat{A}$ is the adjacency matrix (normalized and containing self-loops),
    \item $W^{l}$ and $B^{l}$ are learnable weight and bias terms, and
    \item $\sigma(\cdot)$ is a non-linear activation (ReLU).
\end{itemize}
Each node’s representation is updated by aggregating information from its neighbors. A more explicit illustration of the update rule at the node level can be written as:
\begin{equation}
    h_{i}^{l} 
    = \sigma \left(
        \sum_{j \in N(i)} 
            \frac{1}{|N(i)|} \, W^{l} \, h_{j}^{l-1}
      \right),
\end{equation}
where $N(i)$ is the set of neighbors of node $i$. This formulation underscores that each node embedding $h_{i}^l$ is derived from the aggregation of its neighbors' embeddings at the previous layer.

\paragraph{Normalization.}
In practice, we use degree-based normalization to stabilize training and to ensure consistent scaling across different parts of the graph. This is particularly beneficial when sub-regions (i.e., nodes) vary significantly in their degrees.

\paragraph{Training Objective.}
We train the GNN in a self-supervised manner to reconstruct the adjacency matrix $A$ of the mobility network. After the second GNN layer, we obtain the final node embedding matrix $H^{2}$. We then apply a suitable decoder as a reconstruction function ( additional feedforward layer) to predict $\hat{A}$. The reconstruction loss, Mean Squared Error (MSE), is:
\begin{equation}
    \mathcal{L} 
    = \frac{1}{N^{2}} \sum_{i=1}^{N} \sum_{j=1}^{N} 
      \bigl(A_{ij} - \hat{A}_{ij}\bigr)^{2}.
\end{equation}
By minimizing this loss, the GNN learns node embeddings that capture both global and local connectivity patterns in the graph.

\subsubsection*{MLP-Derived Embedding (Baseline)}

For comparison, we also train a simpler Multi-Layer Perceptron (MLP)---referred to here as a Vanilla Neural Network (VNN)---to learn node embeddings that can reconstruct the adjacency matrix. This baseline method initializes each node with a $d$-dimensional trainable embedding and transforms pairwise concatenations of these embeddings through a feedforward network to predict edge values. Formally, the process involves:

\begin{enumerate}
    \item \textbf{Initialization:} Each node $i$ has an embedding $e_i \in \mathbb{R}^d$.
    \item \textbf{Pairwise Interactions:} For all node pairs $(i, j)$, we concatenate learnable embedding vectors $e_i$ and $e_j$  to form an input feature vector.
    \item \textbf{MLP:} A 3-layer feedforward network processes each pairwise feature vector to yield $\hat{A}_{ij}$.
    \item \textbf{Training:} MSE loss is minimized on the adjacency reconstruction, $\sum_{i,j} (A_{ij} - \hat{A}_{ij})^2$.
\end{enumerate}

While this MLP-based baseline can learn embeddings by directly modeling pairwise relationships, it does not leverage graph convolution or neighbor aggregation, potentially limiting its ability to capture higher-order structural properties in the mobility network. Nonetheless, it serves as a computationally straightforward point of comparison and can sometimes provide surprising effectiveness, especially when the graph is not too large or has strong pairwise signals.

In the following sections, we present experimental results comparing these approaches in terms of community detection and further socioeconomic analysis of the communities. Our primary focus lies in understanding how well graph-based embeddings capture the network communities relative to community detection methods, and also how these communities differ from each other. We evaluate the resulting communities with PSE, Graph embeddings, and community detection methods in 12 cities under consideration. The PSE for all mobility networks were retrieved choosing appropriate embedding dimensionality for different metro regions. Graph representation learning models (VNN, GNN) were trained for 500 epochs with logMSE objective in reconstructing their respective adjacency matrices. The community detection method was run multiple times to ensure stability.

\section*{Results \& Discussion}

\subsection*{Community structures}

\paragraph{Borough/County comparison with clusters}
One challenge in evaluating the detected communities is the absence of a universally recognized "ground truth" for urban delineation. Nevertheless, in some major cities such as New York--comparisons can be drawn against well-established administrative units like boroughs or counties within the city. Since our analysis is performed at the census tract level, we can assess how closely the discovered communities align with these larger administrative boundaries, providing a meaningful benchmark for validating the coherence of our results.

\begin{figure}[ht]
  \centering
  \begin{subfigure}[b]{0.38\linewidth}
    \includegraphics[width=\linewidth]{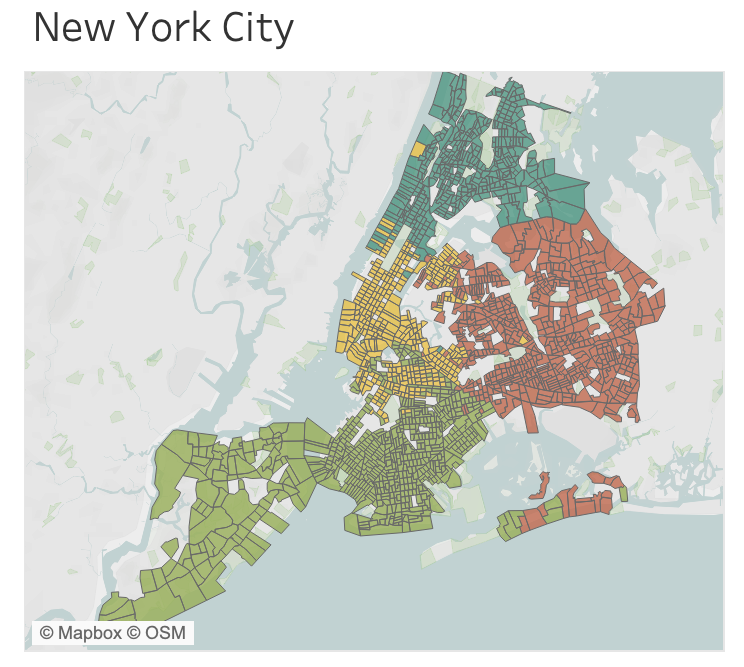}
    \caption{GNN-based embedding communities}
    \label{fig:nyc_gnn}
  \end{subfigure}
  \begin{subfigure}[b]{0.38\linewidth}
    \includegraphics[width=\linewidth]{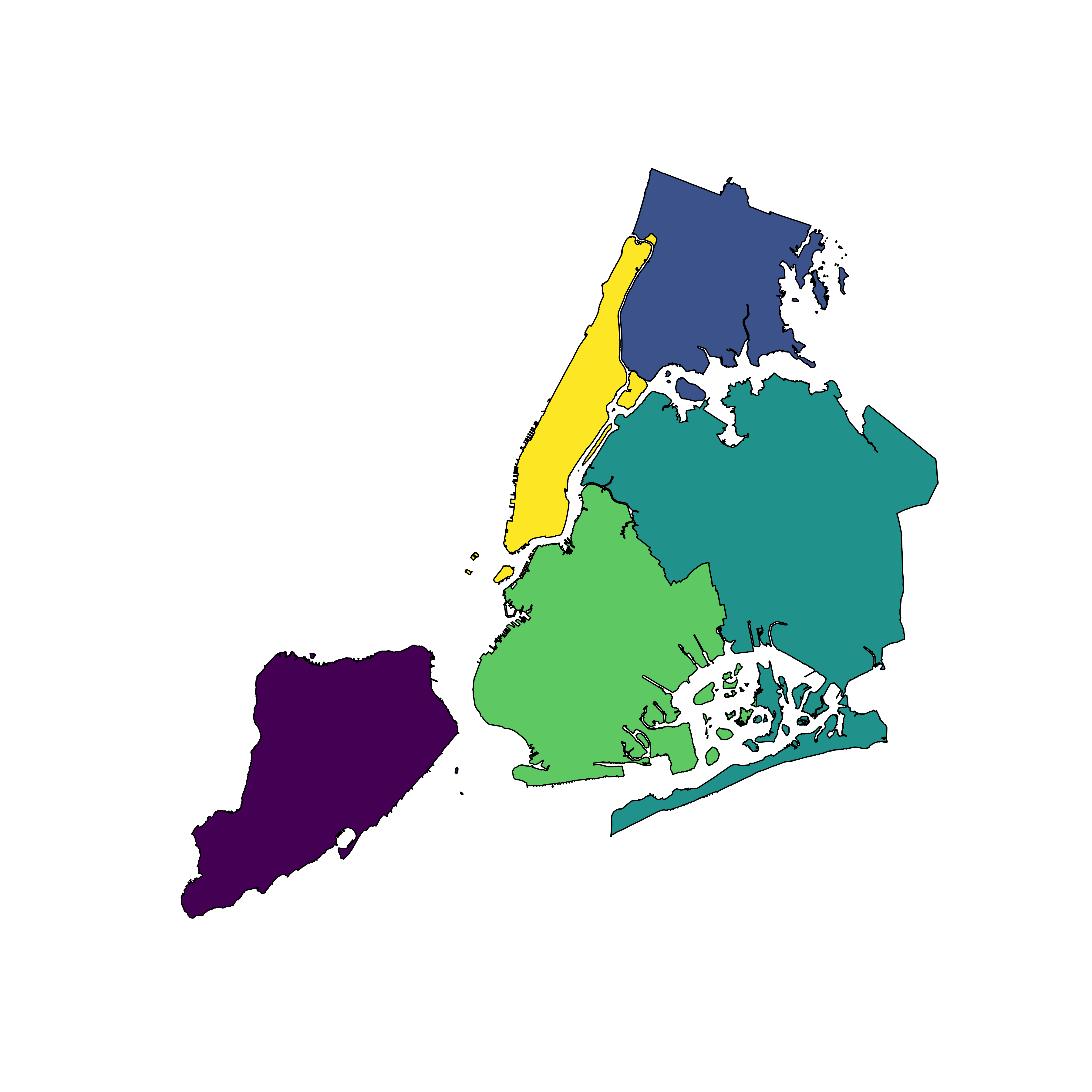}
    \caption{NYC's official borough boundaries}
    \label{fig:chicago_gnn}
  \end{subfigure}
  \caption{Comparison of GNN communities and borough boundaries in NYC}
  \label{fig:community_boroughs}
\end{figure}

An inspection of the GNN-embedding derived communities in Figure~\ref{fig:community_boroughs} reveals that community clusters do not strictly follow the official borough boundaries. For instance, certain areas of southwestern Brooklyn are grouped together with Staten Island, forming a single community despite belonging to different boroughs. Likewise, Manhattan appears partitioned into multiple clusters, indicating diverse mobility and interaction patterns within what is officially one borough. Queens also shows distinct subdivisions rather than forming a single unified cluster. These overlaps and divergences underscore that the detected communities are driven more by patterns in the underlying data (e.g., mobility interaction strengths) than by administrative lines, highlighting both the utility and the potential complexity of such data-driven delineations.

In the subsequent section, we therefore investigate how these communities diverge across key indicators, focusing particularly on socioeconomic metrics, to illuminate the underlying factors that shape these data-driven partitions.

\begin{figure}[H]
  \centering
  \begin{subfigure}[b]{\textwidth}
    \centering
    \begin{subfigure}[b]{0.3\linewidth}
      \includegraphics[width=\linewidth]{img/New_York_City_GNN.png}
    \end{subfigure}
    \hfill
    \begin{subfigure}[b]{0.3\linewidth}
      \includegraphics[width=\linewidth]{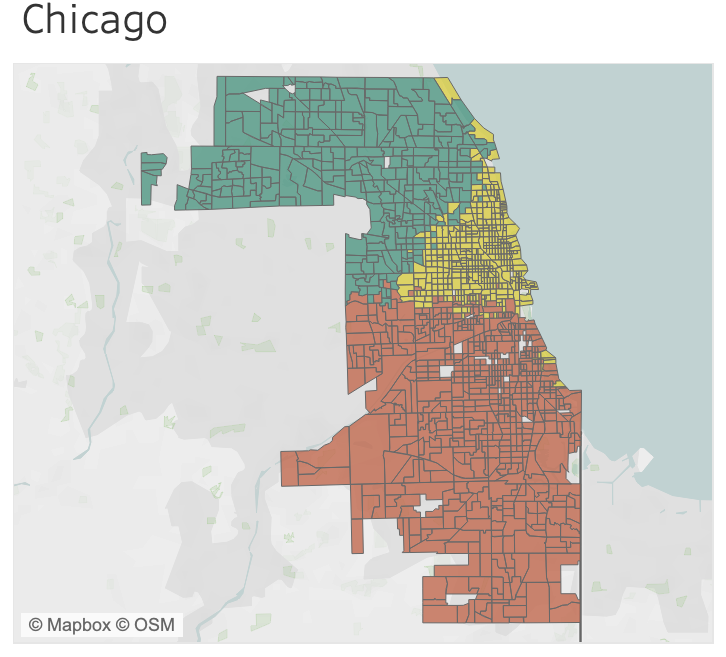}
    \end{subfigure}
    \hfill
    \begin{subfigure}[b]{0.3\linewidth}
      \includegraphics[width=\linewidth]{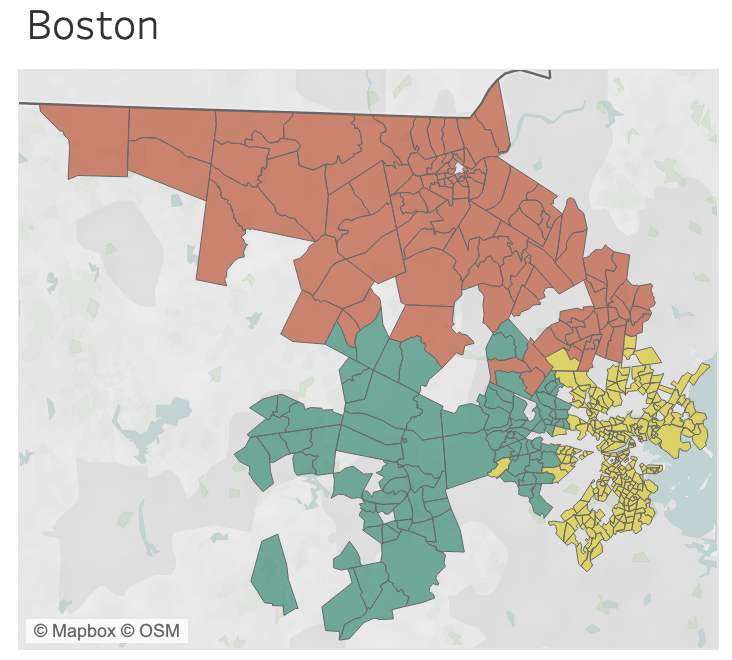}
    \end{subfigure}
    \caption{GNN}
  \end{subfigure}


  \begin{subfigure}[b]{\textwidth}
    \centering
    \begin{subfigure}[b]{0.29\linewidth}
      \includegraphics[width=\linewidth]{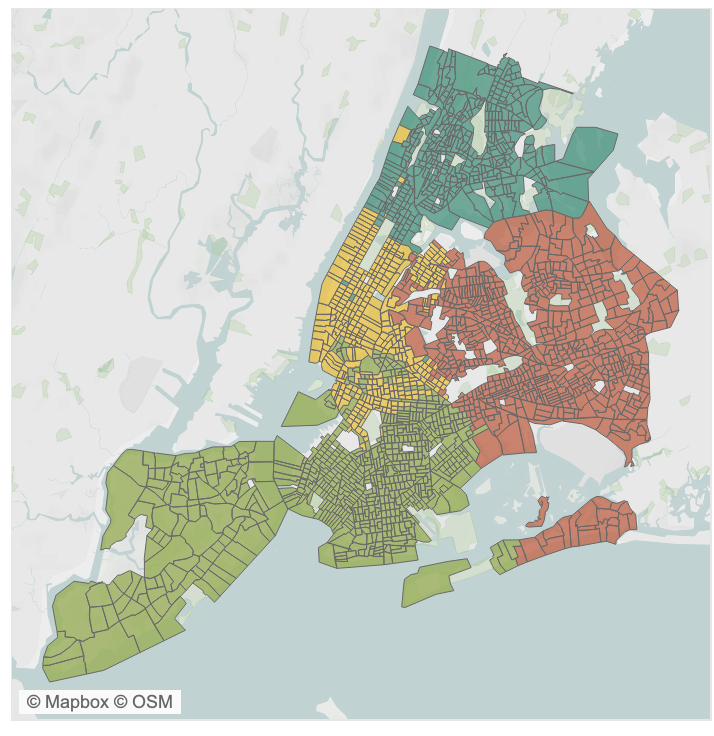}
    \end{subfigure}
    \hfill
    \begin{subfigure}[b]{0.3\linewidth}
      \includegraphics[width=\linewidth]{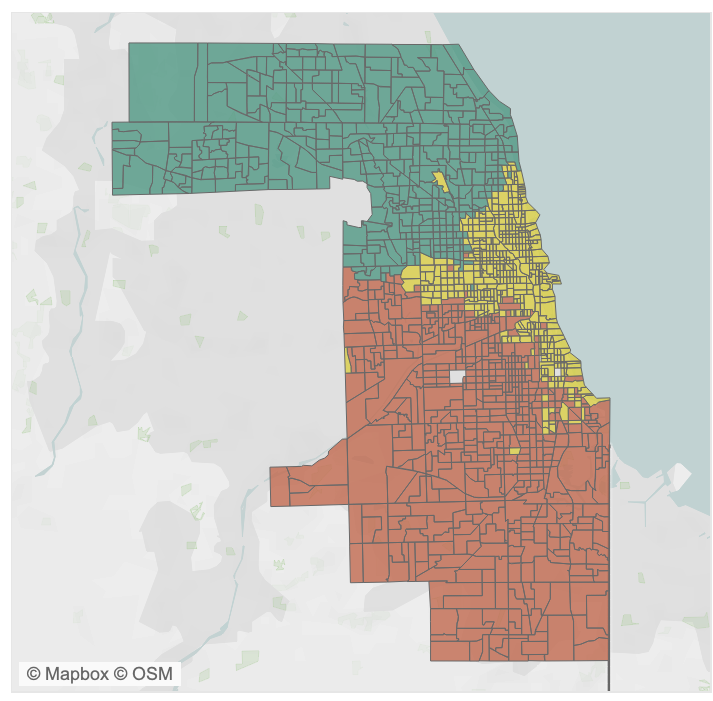}
    \end{subfigure}
    \hfill
    \begin{subfigure}[b]{0.29\linewidth}
      \includegraphics[width=\linewidth]{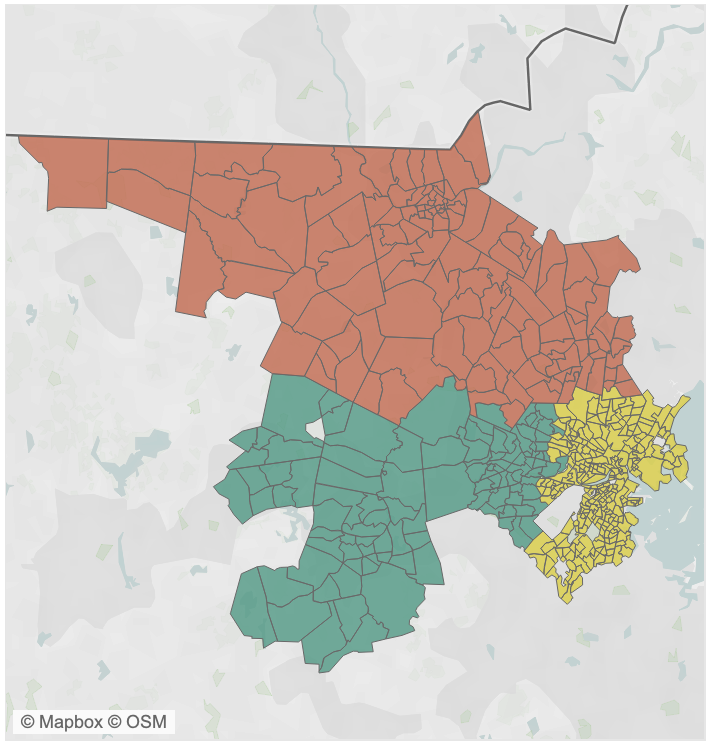}
    \end{subfigure}
    \caption{COMBO}
  \end{subfigure}


    \begin{subfigure}[b]{\textwidth}
    \centering
    \begin{subfigure}[b]{0.3\linewidth}
      \includegraphics[width=\linewidth]{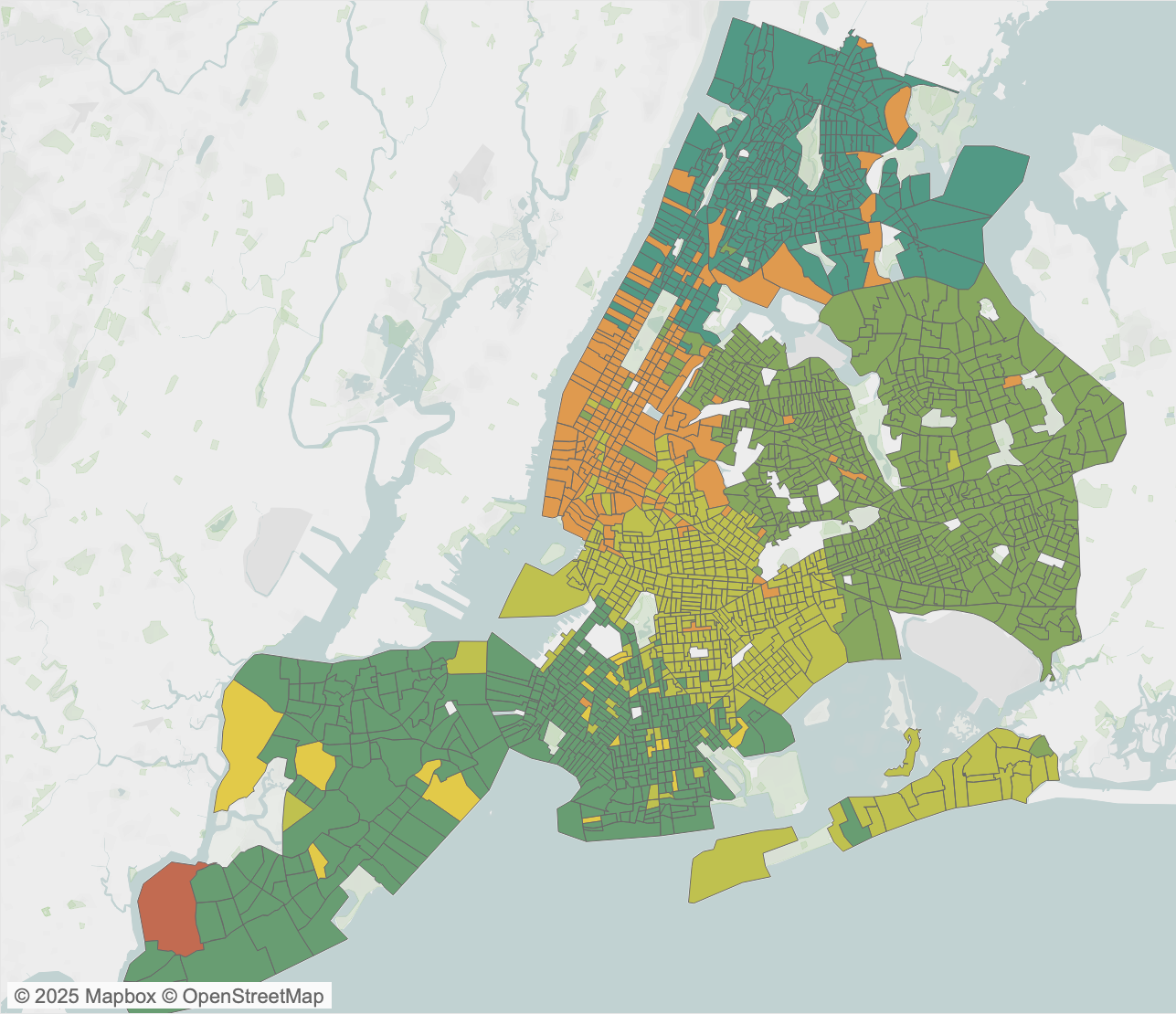}
    \end{subfigure}
    \hfill
    \begin{subfigure}[b]{0.28\linewidth}
      \includegraphics[width=\linewidth]{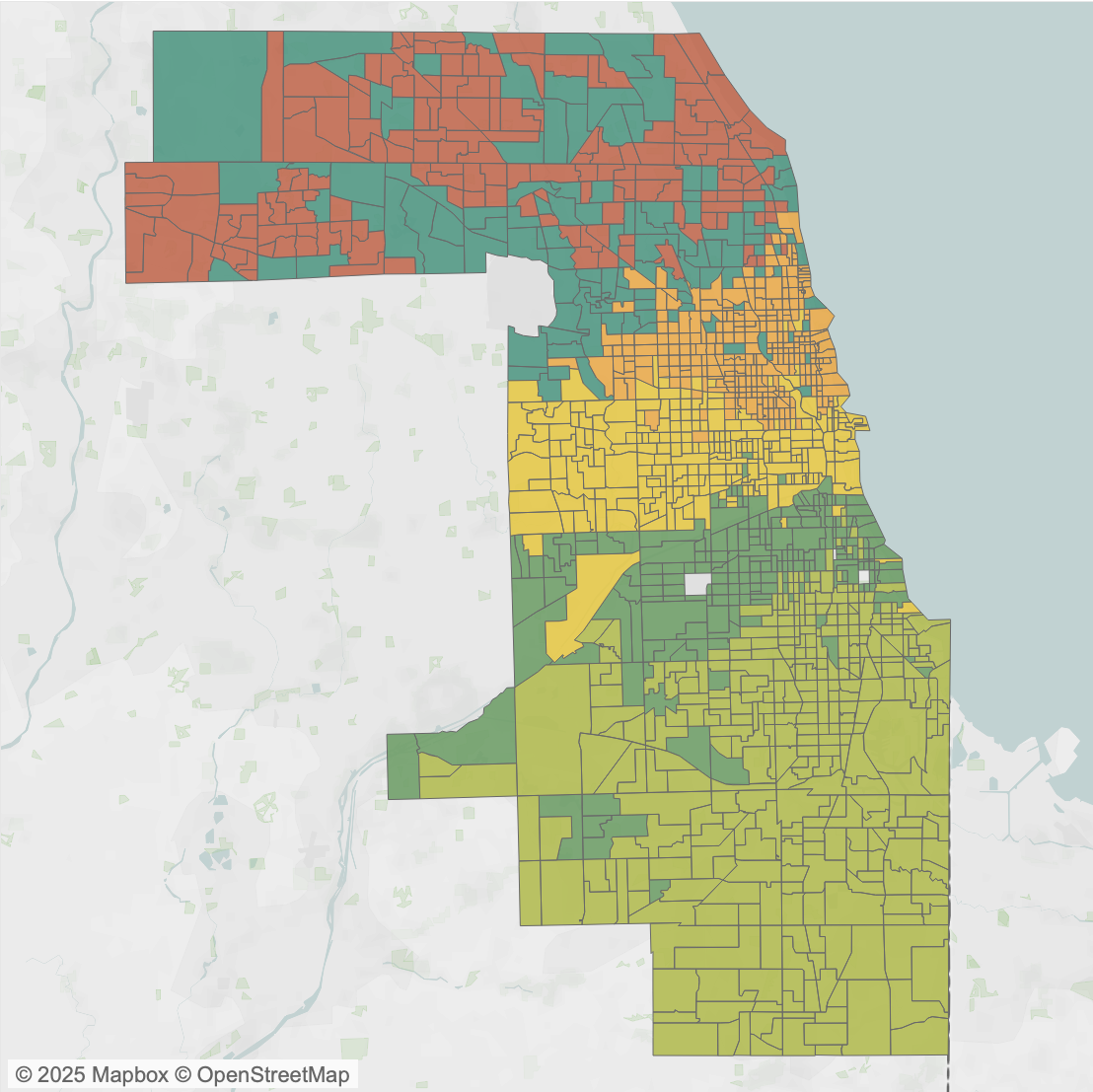}
    \end{subfigure}
    \hfill
    \begin{subfigure}[b]{0.3\linewidth}
      \includegraphics[width=\linewidth]{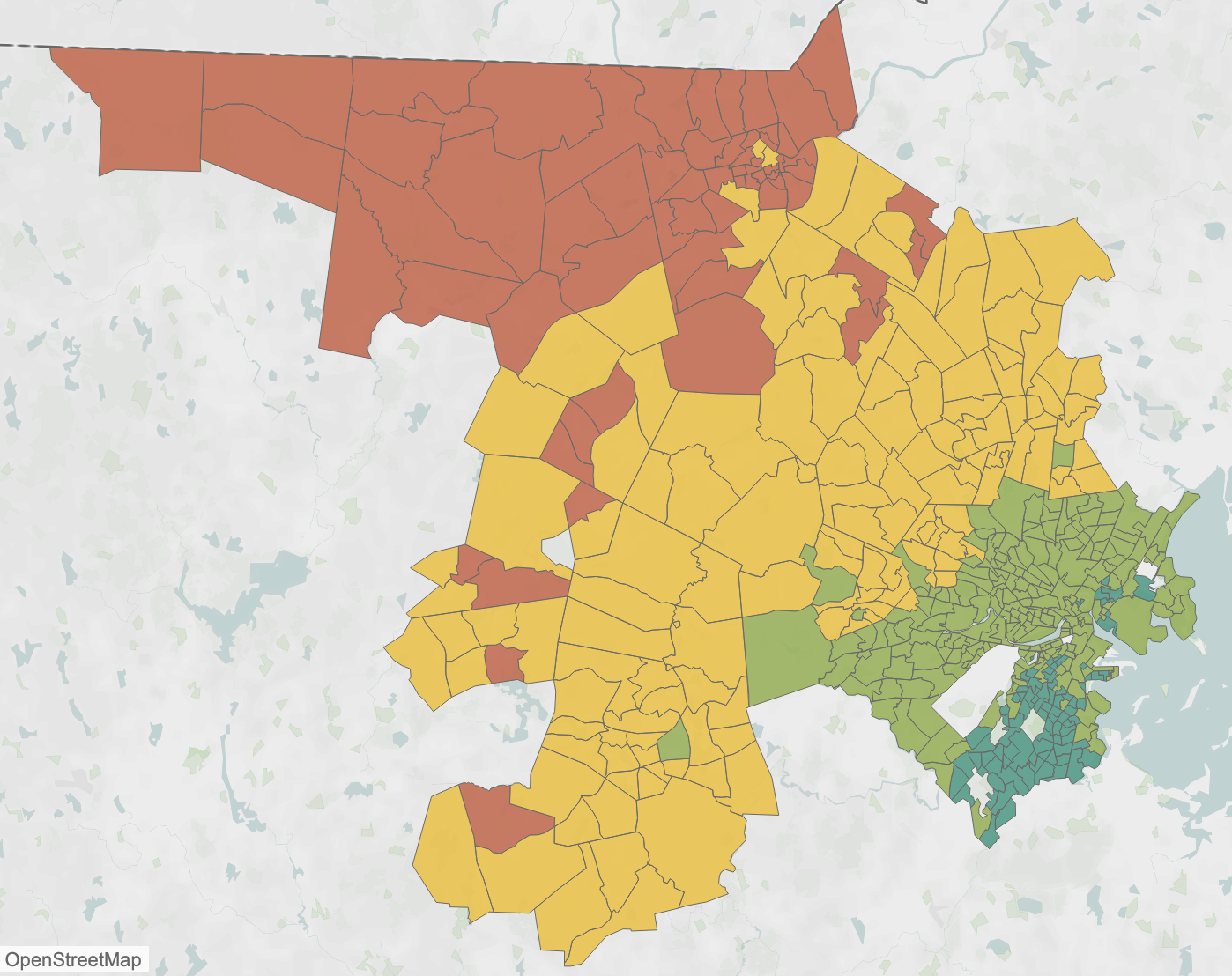}
    \end{subfigure}
    \caption{SBM}
  \end{subfigure}

  \caption{Community structures resulting from 1. GNN-embedding based communities, 2. COMBO method \cite{combo} (bottom row), and 3. Stochastic Block Model (SBM). The GNN-derived community boundaries are similarly aligned to modularity optimization based COMBO. Notably, SBM's results have more communities in all cities. Moreover, SBM-derived community structures are observed to be smaller and not spatially cohesive.}
  \label{fig:community_comparisons}
\end{figure}

\paragraph{Comparisons with community detection methods}
Community detection methods have been commonly used to delineate districts and find urban communities in a variety of networks. Since there is no "ground-truth" as such in terms of urban areas, we can still compare the communities derived from network embeddings to commonly used community detection methods from the literature. 

Traditionally, community detection methods have focused on optimizing modularity, a heuristic measure that quantifies the density of links inside communities compared to links between them \cite{ModularityCommD}. More recent methods have addressed community detection from a principled statistical perspective using Stochastic Block Models (SBMs). SBMs are generative models that assume the observed network structure arises from an underlying block structure, where nodes are assigned to latent groups (or blocks), and edges are generated according to probabilities that depend solely on the group memberships of the nodes \cite{SBMreview}. In practice, inference is performed by minimizing the description length of the model using Bayesian model selection techniques, such as the Minimum Description Length (MDL) principle. 

Fig~\ref{fig:community_comparisons} also shows a comparison with two distinct and well-established approaches for community detection in networks. We evaluate and compare the communities obtained from 1. COMBO method \cite{combo}, a well-established benchmark for community detection in networks based on modularity optimization, and 2. An SBM model based on \cite{SBMmcmc} with Monte Carlo optimization. In all cities in our analysis, we notice GNN-embedding based communities closely align with those we get from COMBO method. This is also quantitatively shown by evaluating the modularity for communities from both methods. The community delineations produced by the SBM approach differ notably from those of other methods, often resulting in smaller and spatially incohesive communities in most cities. In New York City, for example, while SBM successfully identified neighborhoods in Manhattan and the Bronx (similar to GNN and COMBO), it also isolated a single census tract as its own community, highlighting its tendency toward fragmented partitions. In Chicago, SBM's results diverged more significantly, yielding scattered and irregularly shaped communities that contrast sharply with the more coherent patterns detected by GNN and COMBO. A similar trend was observed in Boston and Chicago, where the communities derived from SBM were consistently smaller and structurally different from those produced by the other two methods. Interestingly, for Austin and San Antonio, SBM assigns all nodes of the network to a single community. This contrasts GNN and COMBO methods, which have 3 cohesive communities for both cities.

Table~\ref{tab:city_NMIstats} presents the modularity scores for the communities identified in each city. Across the 12 cities analyzed, we find that communities derived from GNN-based embeddings achieve modularity scores comparable to those produced by COMBO. In contrast, the spatial fragmentation of SBM-derived communities results in substantially lower modularity scores. Similarly, communities obtained from MLP and PSE-based embeddings also lack spatial coherence and correspondingly exhibit significantly reduced modularity.


\hfill
\begin{table}[h]
  \caption{Modularity scores for communities resulting from different embeddings compared with COMBO-based network partitions}
  \label{tab:city_NMIstats}
  \begin{tabular}{lccccc} 
    \textbf{City} & \textbf{Number of communities} & \textbf{COMBO} & \textbf{SBM} & \textbf{GNN} & \textbf{VNN baseline}\\
    New York & 4 & 0.27 & 0.16 & 0.259 & 0.12 \\
    Chicago & 3 & 0.275 & 0.13 & 0.264 & 0.10  \\
    Boston & 3 & 0.272 & 0.14 & 0.259 & 0.16   \\
    Austin & 3 & 0.162 & 0.0 & 0.156 & 0.07  \\
    Dallas & 3 & 0.187 & 0.08 & 0.187 & 0.05 \\
    Los Angeles & 4 & 0.354 & 0.19 & 0.351 & 0.18  \\
    San Antonio & 3 & 0.169 & 0.0 & 0.168 & 0.07 \\
    San Diego & 4 & 0.30 & 0.18 & 0.29 & 0.17 \\
    San Jose & 4 & 0.196 & 0.10 & 0.188 & 0.11 \\
    Philadelphia & 4 & 0.185 & 0.12 & 0.179 & 0.09 \\
    Phoenix & 3 & 0.24 & 0.14 & 0.24 & 0.13 \\
    Houston & 3 & 0.25 & 0.18 & 0.25 & 0.07 \\
  \end{tabular}
\end{table}


\subsection*{Socioeconomic evaluation}

\begin{figure}[H]
  \centering
  \begin{subfigure}[b]{0.38\linewidth}
    \includegraphics[width=\linewidth]{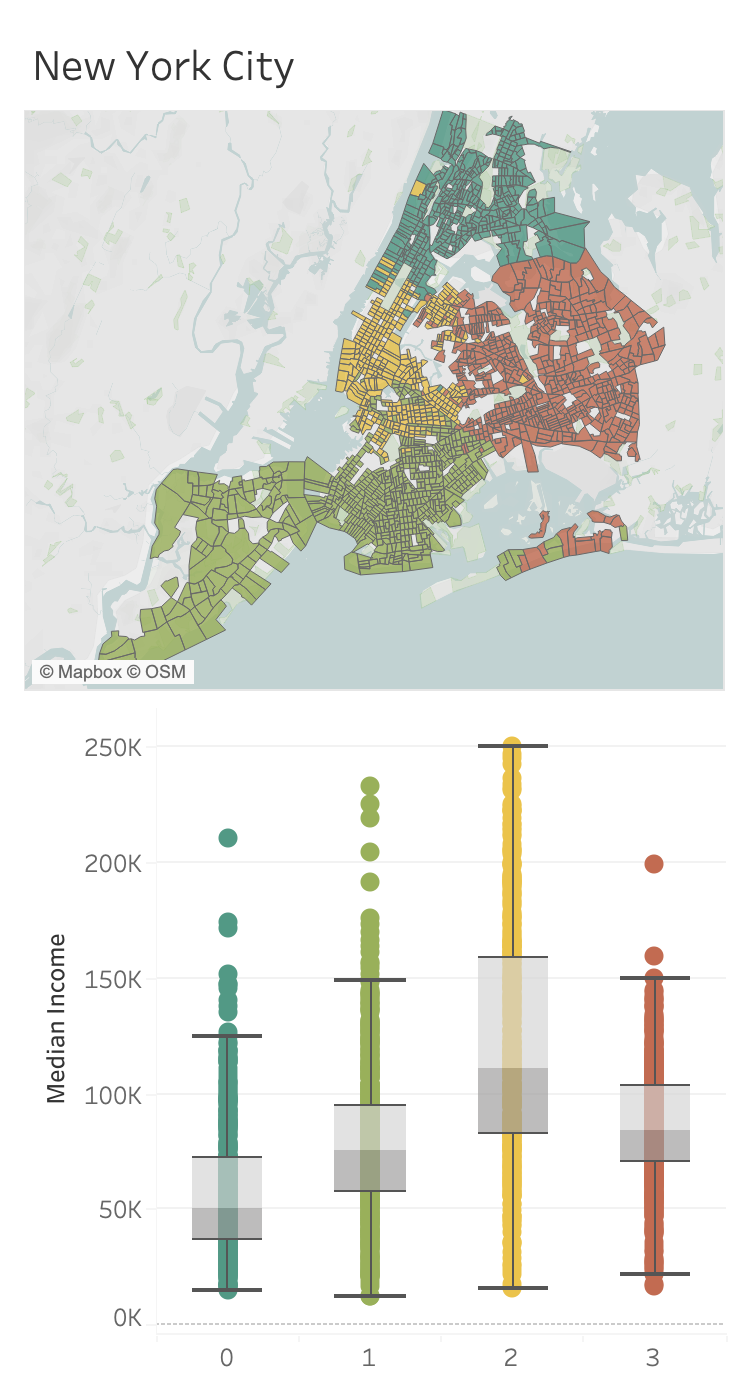}
    \caption{New York City}
    \label{fig:nyc_gnn}
  \end{subfigure}
  \begin{subfigure}[b]{0.38\linewidth}
    \includegraphics[width=\linewidth]{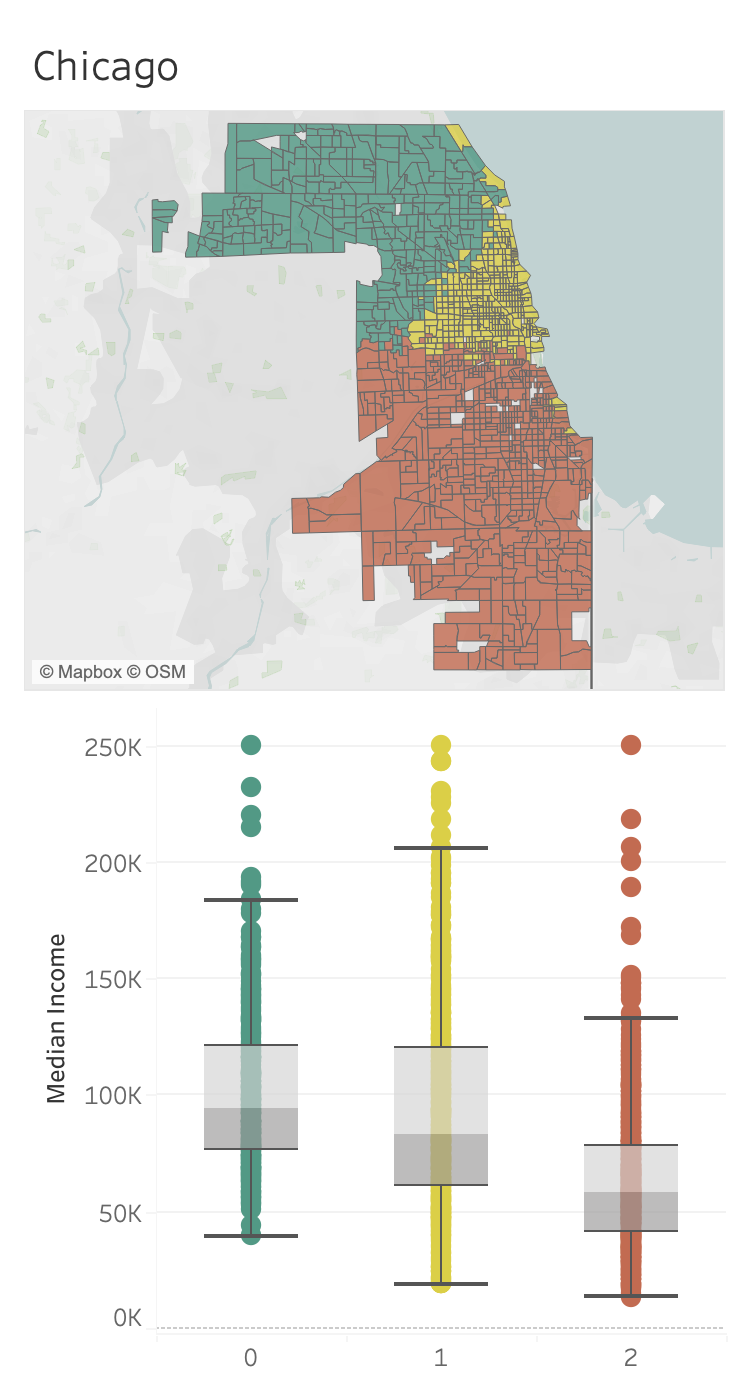}
    \caption{Chicago}
    \label{fig:chicago_gnn}
  \end{subfigure}
  \caption{GNN-embedding based communities -- shown along with income distributions within each community.}
  \label{fig:community_income_dist}
\end{figure}

To further investigate inter-cluster differences, we analyze the distribution of a key socioeconomic indicator--median neighborhood income--across the identified communities. Median income is a widely recognized proxy for residents’ socioeconomic status and provides a meaningful basis for comparison. Our analysis shows that the communities delineated using network embeddings—particularly those generated by GNN-based methods—exhibit distinct income profiles. As shown in Fig.~\ref{fig:community_income_dist}, we analyze the income distributions of census tracts and examine the resulting disparities across all communities. While not all communities are sharply differentiated, there is a clear separation between high- and low-income areas, with many of them forming distinct and internally consistent clusters.

\begin{figure}[H]
    \centering
    \begin{subfigure}[b]{0.4\textwidth} 
        \centering
        \includegraphics[width=\textwidth]{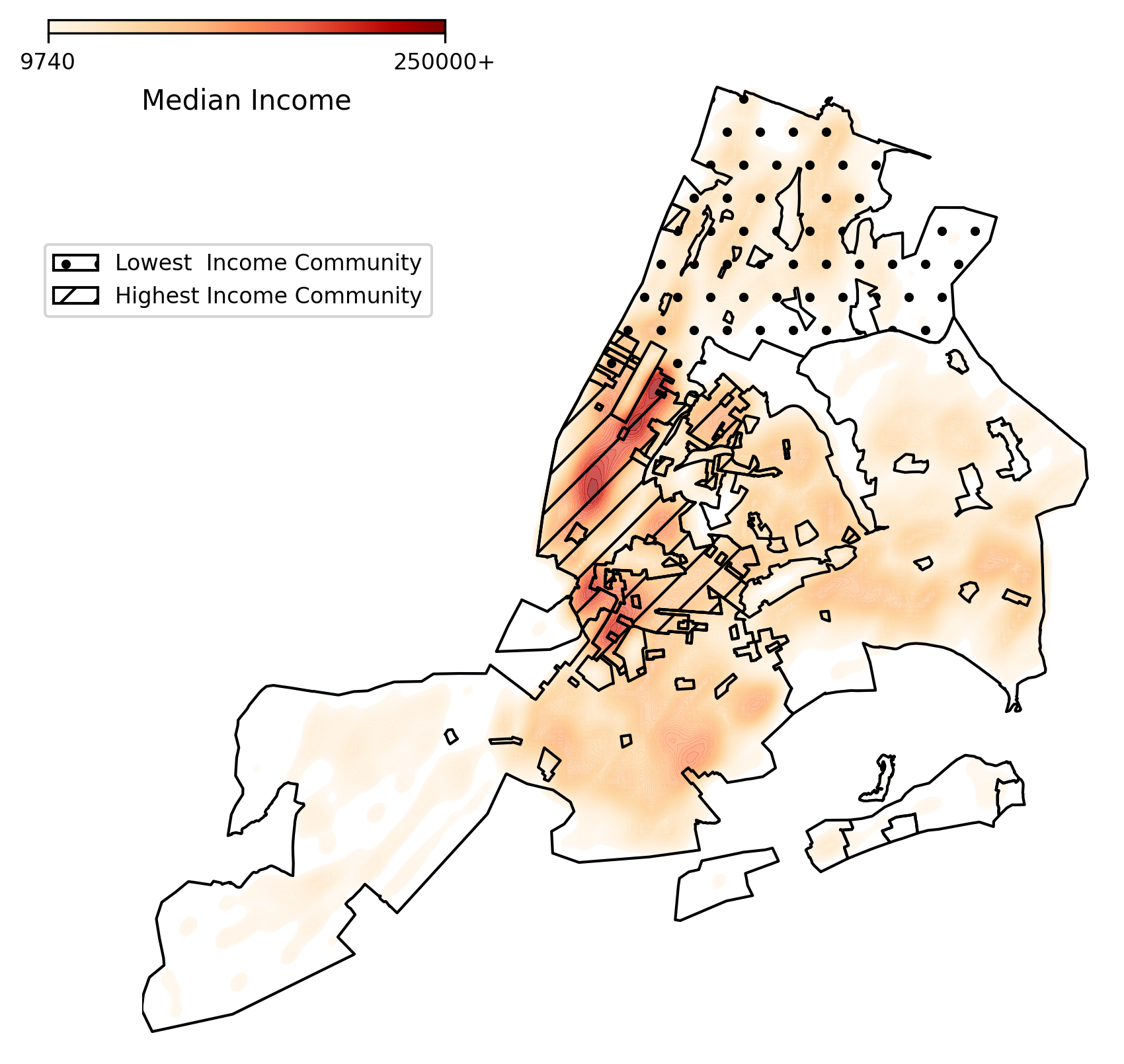} 
        \caption{New York}
        \label{fig:subfigure1}
    \end{subfigure}
    \hfill
    \vspace{1em}
    \begin{subfigure}[b]{0.4\textwidth} 
        \centering
        \includegraphics[width=\textwidth]{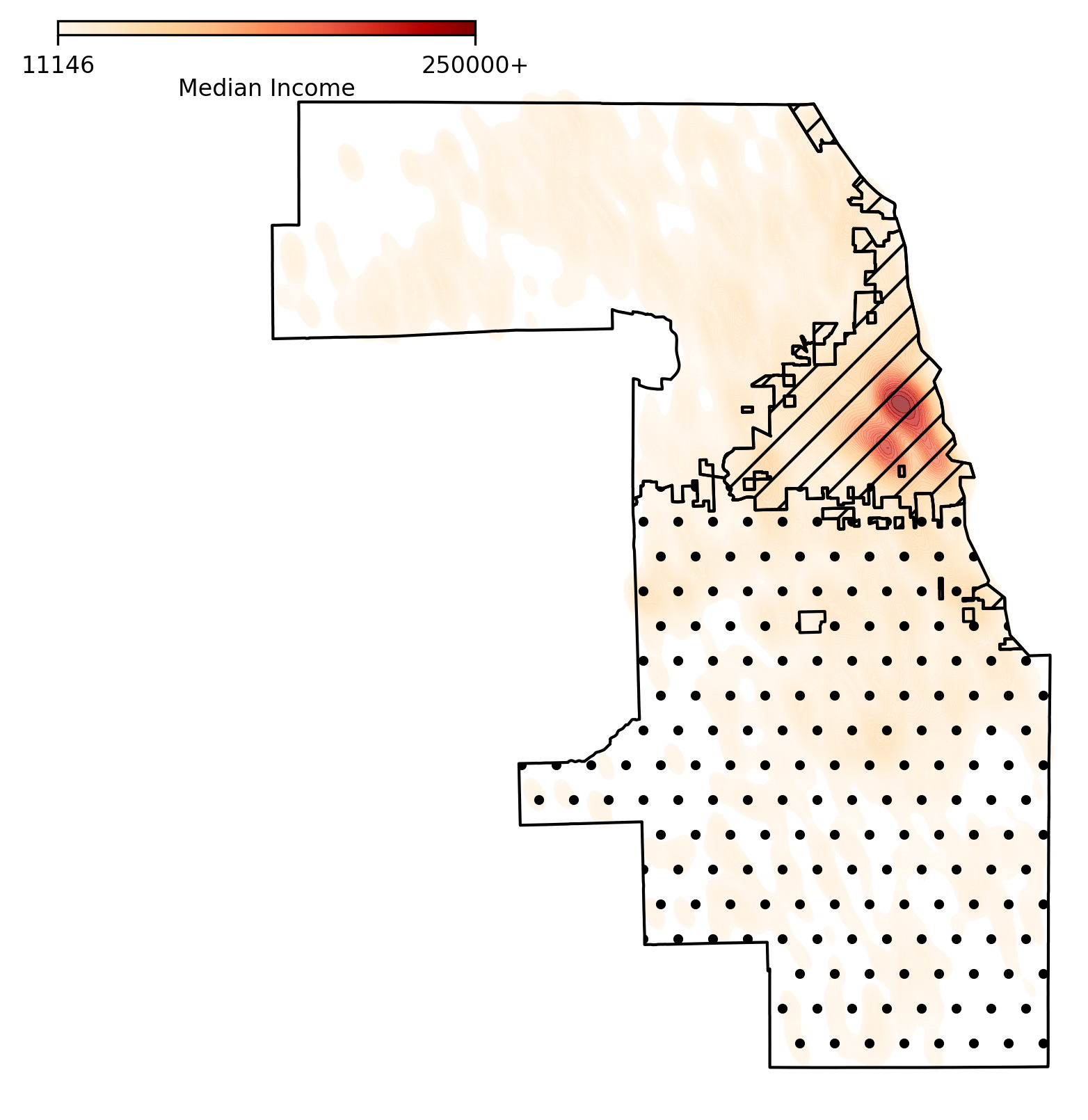} 
        \caption{Chicago}
        \label{fig:subfigure2}
    \end{subfigure}
    \hfill
    \vspace{1em}
    \begin{subfigure}[b]{0.4\textwidth} 
        \centering
        \includegraphics[width=\textwidth]{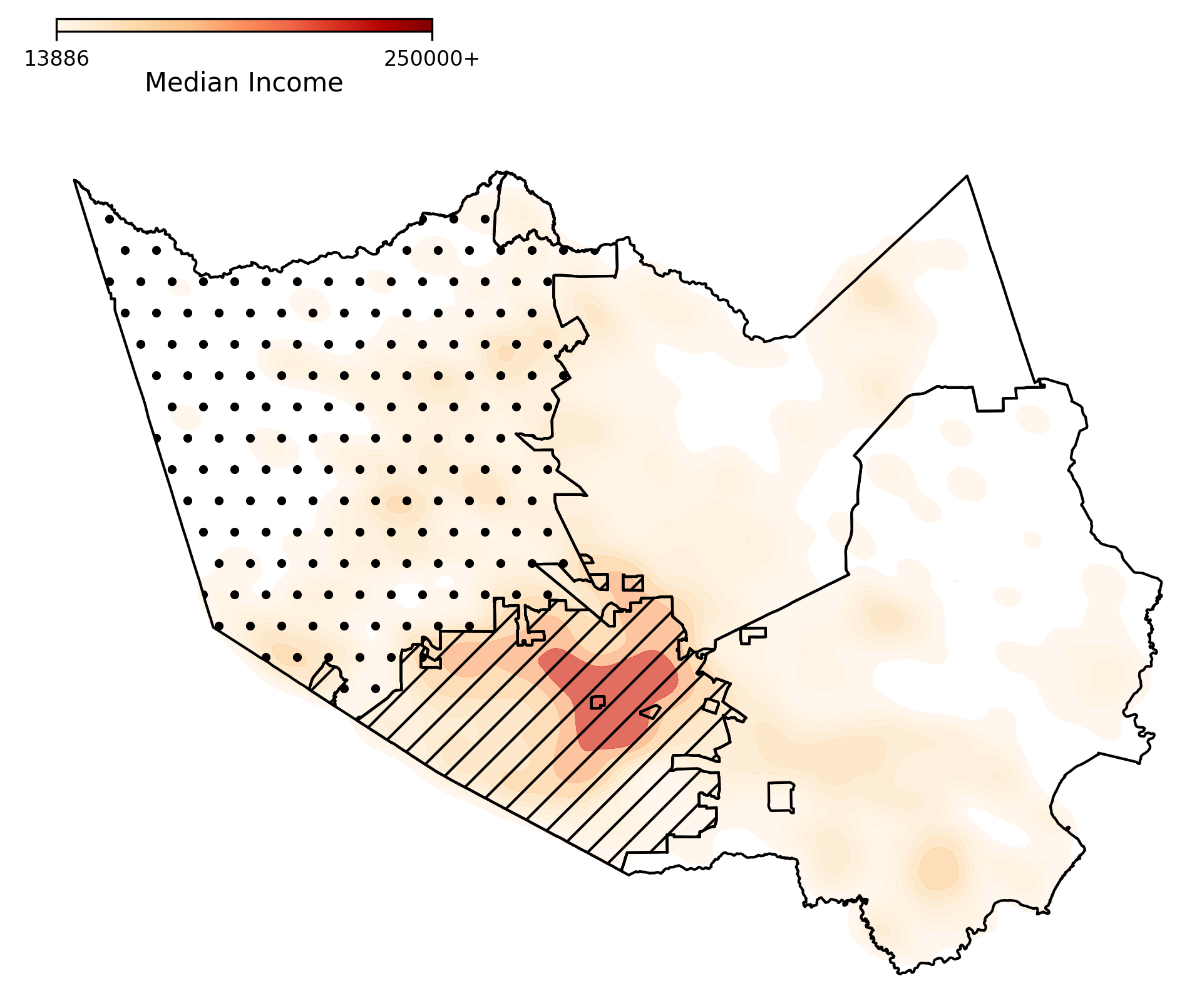} 
        \caption{Houston}
        \label{fig:subfigure2}
    \end{subfigure}
    \hfill
    \vspace{1em}
    \begin{subfigure}[b]{0.4\textwidth} 
        \centering
        \includegraphics[width=\textwidth]{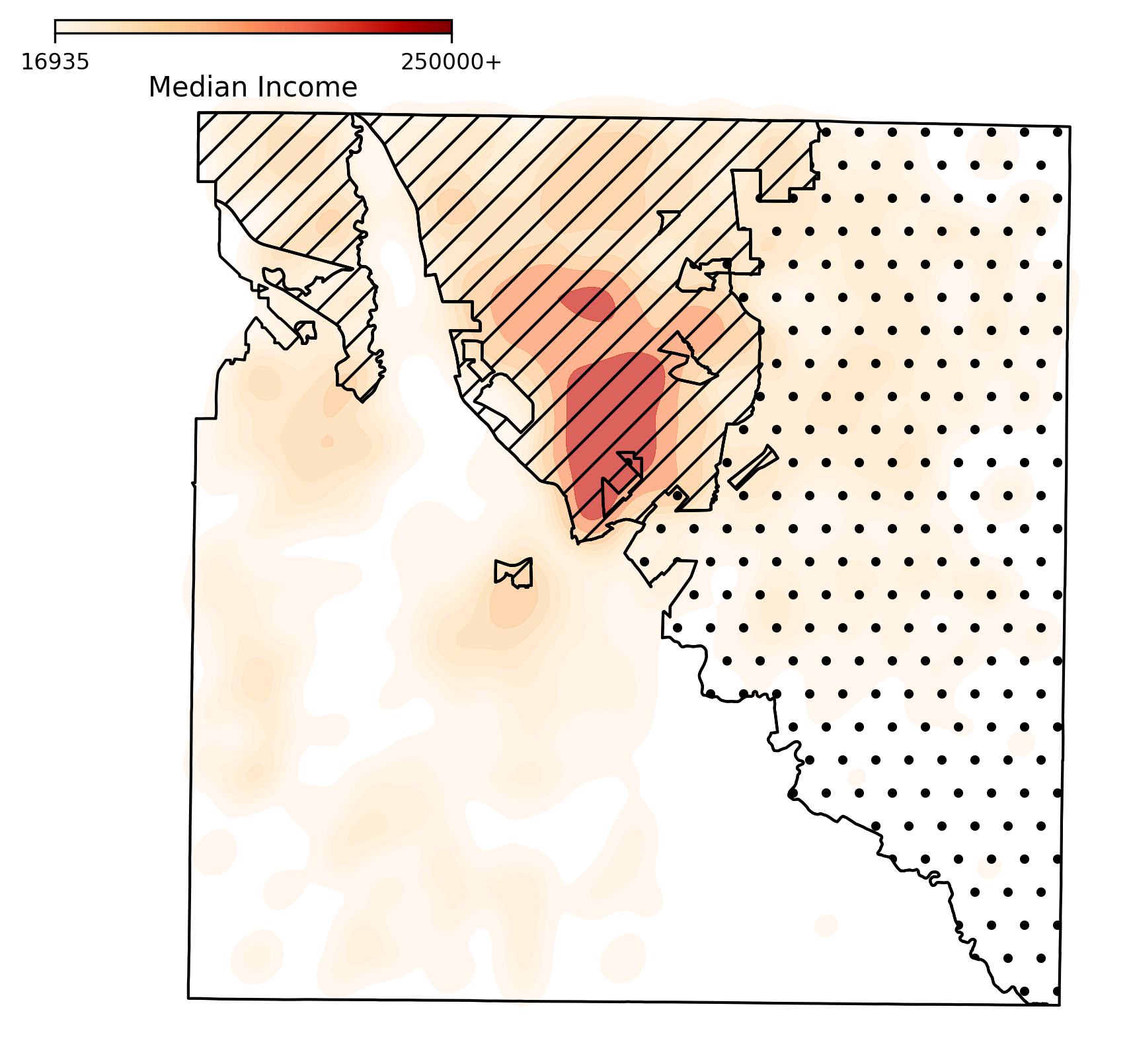} 
        \caption{Dallas}
        \label{fig:subfigure2}
    \end{subfigure}
    \hfill
    \vspace{1em}
    \begin{subfigure}[b]{0.4\textwidth} 
        \centering
        \includegraphics[width=\textwidth]{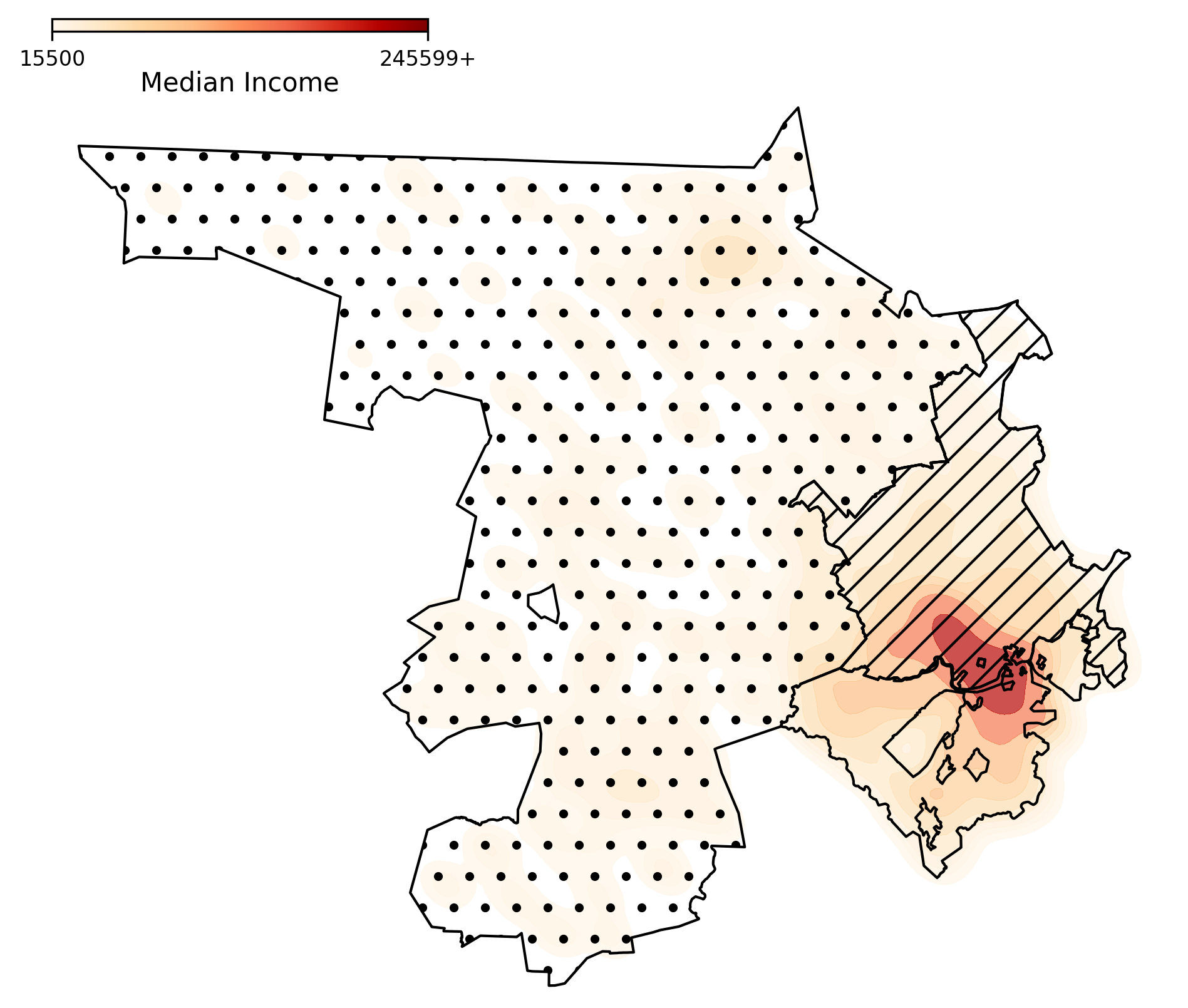} 
        \caption{Boston}
        \label{fig:subfigure2}
    \end{subfigure}
    \hfill
    \vspace{1em}
    \begin{subfigure}[b]{0.4\textwidth} 
        \centering
        \includegraphics[width=\textwidth]{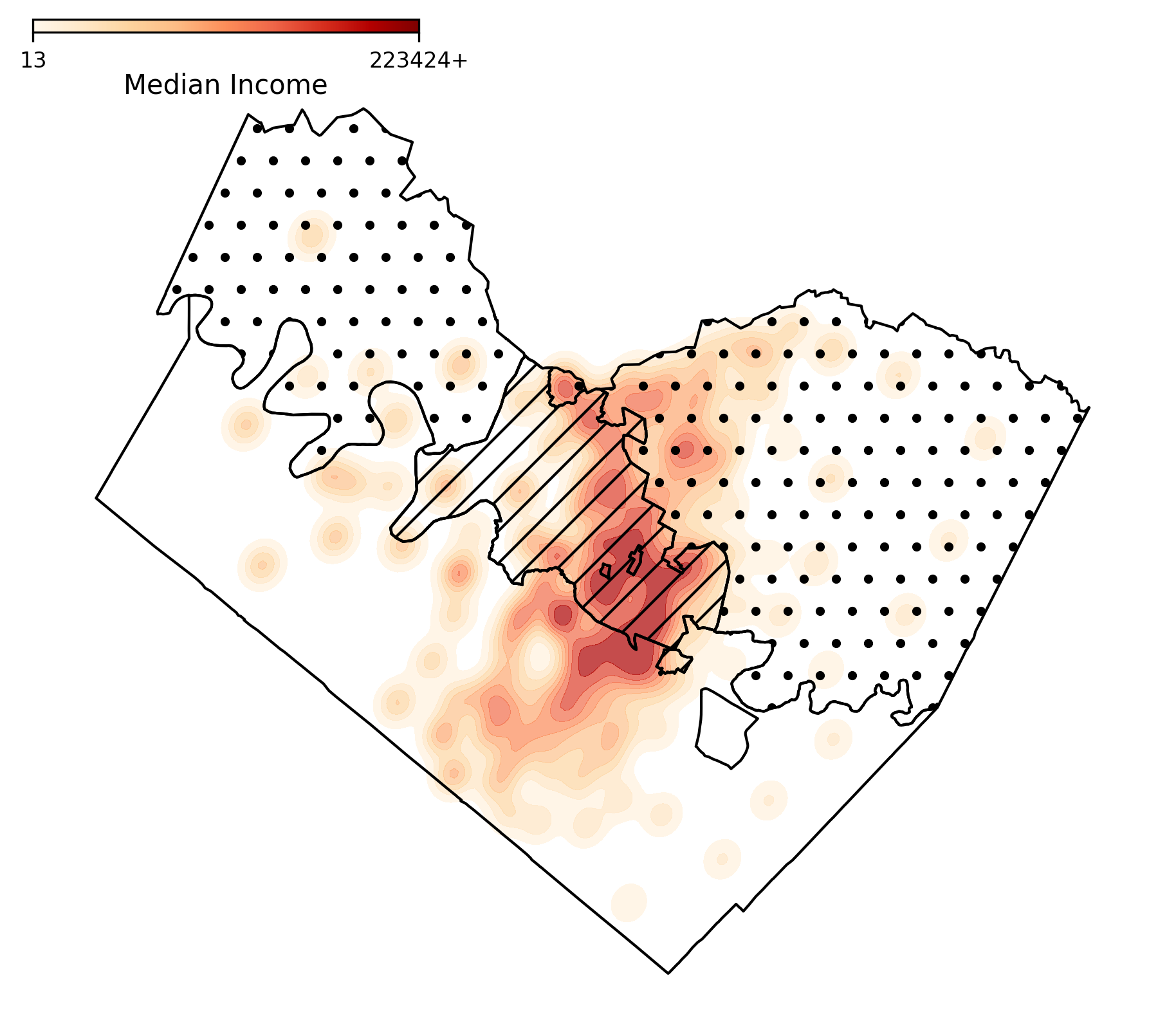} 
        \caption{Austin}
        \label{fig:subfigure2}
    \end{subfigure}
    \hfill
    \vspace{1em}
    \caption{Communities are distinguished by their socioeconomic status -- highest income density areas are captured by a community in most cities. Some cities -- like Boston, Austin, and Dallas-- have high-income areas lying in multiple different communities. Generally, the delineated communities capturing wealthier income groups tend to be smaller compared to other communities in all cities.}
    \label{fig:high-low-communities}
\end{figure}

Across several focal cities, community‐detection results align closely with the spatial distribution of income density. As shown in Fig.~\ref{fig:high-low-communities} in New York, for instance, the core of Manhattan emerges as one or two distinct high‐income communities that sharply contrast with outlying, lower‐income clusters. In Chicago, higher‐income density is concentrated primarily along the northern corridor, with that region split between two overlapping communities--both capturing slightly different slices of affluence. Some cities reveals a larger, multi‐community overlap surrounding high‐income enclaves. Boston’s high‐income center is more tightly contained; however it is spanned by two communities enclosing the downtown area. By contrast, in Houston and Phoenix, wealthier zones sprawl outward from the central core. Dallas also exhibits a substantial high‐income cluster largely confined to one community, but with a few additional pockets identified by neighboring communities. These individual cases demonstrate how community detection can trace the contours of both concentrated and dispersed affluence across different urban environments. Since our analysis is based on commute mobility data, the spatial clustering of high-income neighborhoods within a single community indicates that individuals from higher income groups tend to reside and travel within the same localized areas. Notably, a comparison of community sizes reveals that high-income communities are generally smaller across all cities, suggesting that affluent populations exhibit more spatially concentrated and possibly denser mobility patterns. In contrast, low-income communities tend to span larger areas, implying that less affluent individuals are more likely to commute over longer distances. This observation aligns with previous studies that have examined the relationship between income levels and commuting behavior \cite{IncomeCommuteImpact,IncomeCommute2}. The income differences among communities are quantified by Jensen–Shannon (J–S) divergence score, a symmetric measure of similarity between two distributions. This metric, bounded between 0 and 1, effectively captures the degree of disparity in income distributions across communities. Divergence based metrics, including J-S divergence, have been used in the literature to assess differences in income and other economic factors among the population.\cite{kirkley2020JSdivergence, JSdivergence1, magdalou2011income}. It thus provides us with a metric to compare the income distribution among the delineated districts in our analysis. Table~\ref{tab:JS_scores} presents a comparison of high‐ versus low‐income community distributions—as measured by the J–S divergence. It reveals considerable variation in how sharply these two groups diverge across different cities. In places such as Philadelphia and Austin, the high‐income and low‐income communities exhibit noticeably distinct distributional profiles, suggesting stark socioeconomic contrasts. By contrast, Chicago and Los Angeles show more moderate divergence values, implying relatively more overlap—or less pronounced differences—between the characteristics of their highest‐ and lowest‐income communities. Cities like New York and Boston fall somewhere in between: although high‐ and low‐income distributions differ substantially, those differences are not as extreme as in Philadelphia, nor as subtle as in LA. In general, these divergences highlight how, in some urban areas, economic disparities manifest as more distinct communities than in others. 

Interestingly, when comparing the income profiles of resulting communities from COMBO, we notice that our method achieve at least the same, and in some cities, higher J-S divergence scores. This suggest that communities identified through our embedding-driven framework are more sharply demarcated in terms of socioeconomic attributes, likely due to the method's ability to preserve fine-grained node attribute correlations during representation learning. This is particularly seen for major metro cities like NYC, Chicago, and Los Angeles, which are the three biggest in the country. In case of SBM, we see some higher J-S divergence scores for some cities like New York, Chicago, and Los Angeles. This is because the method delineates some very small, highly affluent neighborhoods as a single community. Thus, this increased socioeconomic differentiation from SBM comes at the expense of generating overly fragmented and spatially scattered communities.

Overall, GNN-derived communities across all cities successfully capture clear distinctions between high‐ and low‐income areas.  Interestingly, some cities exhibit a single, contiguous high‐income community that encloses their principal area of elevated median income, whereas others reveal two or more communities collectively spanning these high‐income clusters. This pattern suggests that while a single “hub” of wealthy neighborhoods dominate in certain urban areas, others have multiple pockets of high‐income concentration (for instance in Austin). Importantly, the ability of our approach to isolate these communities--even when they are spatially dispersed--emphasizes its utility in analyzing urban income patterns.

\begin{table}[h]
  \centering
  \caption{J-S divergence scores and median income differences between highest and lowest income communities}
  \label{tab:JS_scores}
  \resizebox{\textwidth}{!}{%
  \begin{tabular}{lcccr}
    \toprule
    \textbf{City} & \multicolumn{3}{c}{\textbf{J-S divergence}} & \textbf{Median income delta (USD)} \\
    \cmidrule(lr){2-4}
    & \textbf{COMBO} & \textbf{SBM} & \textbf{GNN embeddings} & \\
    \midrule
    New York & 0.61 & 0.79 & 0.62 & 58,656\\
    Chicago & 0.49 & 0.79 & 0.54 & 27,793 \\
    Boston & 0.65 & 0.64 & 0.65 & 42,839 \\
    Austin & 0.79 & 0.0 & 0.79 & 67,462  \\
    Dallas & 0.59 & 0.54 & 0.59 & 28,058\\
    Los Angeles & 0.34 & 0.72 & 0.38 & 14,400 \\
    San Antonio & 0.67 & 0.0 & 0.67 & 30,001  \\
    San Diego & 0.66 & 0.65 & 0.66 & 36,630 \\
    San Jose &  0.73 & 0.71 & 0.74 & 46,498 \\
    Philadelphia & 0.76 & 0.75 & 0.76 & 40,440 \\
    Phoenix & 0.48 & 0.43 & 0.48 & 20,834\\
    Houston & 0.47 & 0.47 & 0.47 & 7,657\\
    \bottomrule
  \end{tabular}
  }
\end{table}


\subsection*{Computational advantage}
The adoption of general-purpose graph embeddings for community detection offers substantial computational advantages over traditional methods, particularly when scalability is critical. Conventional approaches, such as the COMBO algorithm, often suffer from high time complexity due to their reliance on iterative pairwise node comparisons or modularity maximization. For instance, executing COMBO on the New York City (NYC) network—a moderately sized graph with around two thousand nodes—requires approximately 4.8 seconds per run on a standard CPU. In contrast, embedding-based methods decouple the computationally intensive graph representation phase from the clustering step. Downstream clustering algorithms (e.g., k-means or spectral clustering) can operate on the embeddings with remarkable efficiency. Clustering the NYC network in the embedded space reduces runtime to just 32 milliseconds, achieving a 150-fold speedup over COMBO. While training a GNN entails an initial computational cost, these embeddings can be reused for multiple analyses without retraining or re-running community detection from scratch. Although raw speedup alone is unlikely to shift practical priorities in urban delineation--where real-time computation is rarely a strict requirement, this framework supports extensibility and efficiency for applied urban analytics, particularly when integrated with other tasks such as prediction, classification, or temporal analysis.

\section*{Conclusion}

This study demonstrates the effectiveness of mobility network embeddings in uncovering urban community structures and revealing their underlying socioeconomic dynamics. While previously established community detection approaches often have a network metric to optimize, our methods show that a general purpose embedding resulting from a self-supervised trained model can provide meaningful urban communities. The analysis across multiple cities shows that the communities detected via GNN-based embeddings capture meaningful patterns that often can diverge from traditional administrative boundaries. For example, while official delineations in cities like New York suggest clear borough divisions, our results indicate that neighborhoods with similar mobility and interaction patterns may span multiple boroughs, suggesting a more nuanced urban fabric.

When benchmarked against established methods, the network embedding approach achieved comparable modularity scores, underscoring its robustness in delineating spatially cohesive clusters. Furthermore, socioeconomic evaluation reveals that data-driven communities align closely with key indicators such as income distribution. The differences in communities among cities vary, with some cities showing sharp delineation in high-income and low-income density areas, whereas in others the differences are more moderate. The variation in income densities among clusters underscores how economic disparities manifest differently depending on local mobility dynamics. This provides valuable insights into the spatial distribution of socioeconomic status based on where people commute within the city.

Another critical insight from our work is the role of commute networks in shaping these urban communities. The flow of commuters and the structure of transit routes play a significant role in connecting disparate neighborhoods, influencing both the formation of community clusters and the observed socioeconomic patterns. By incorporating commute networks into our analysis, we capture a more comprehensive picture of urban mobility, underscoring their importance in the delineation and evolution of community structures. While prior studies have predominantly relied on dynamic mobility data derived from cell phones and social media, our analysis is based on census-based mobility datasets. These datasets offer the distinct advantage of being universally accessible across spatial geographies and demographic groups. Consequently, our findings underscore the value of census-derived mobility data as a critical resource for urban analysis, particularly in contexts where obtaining large-scale mobility data from alternative sources is challenging or cost-prohibitive.

\section*{Acknowledgments}
This research was supported by the MUNI Award in Science and Humanities (MASH Belarus) of the Grant Agency of Masaryk University under the Digital City project~(MUNI/J/0008/2021). This work was also partially supported by the NYUAD Center for Interacting Urban Networks (CITIES), funded by Tamkeen under the NYUAD Research Institute Award CG001.

\section*{Data availability}
The data used in this study can be found in a Zenodo repository: \url{https://doi.org/10.5281/zenodo.11494208}.
Specifically, the data contains origin-destination daily commute flow information among the census tracts in all 12 U.S. cities considered in this work.


\bibliography{references}

\end{document}